# A Novel Symmetric Key Cryptographic Technique at Bit Level Based on Spiral Matrix Concept

Manas Paul[1], Jyotsna Kumar Mandal[2]

[1] Asst. Prof., Dept. of Computer Application, JIS College of Engineering, Kalyani, West Bengal, India
[2] Prof., Dept. of C.S.E., Kalyani University, Kalyani, West Bengal, India
[1] manaspaul@rediffmail.com, [2] jkmandal@rediffmail.com

**ABSTRACT**:- *In this paper, we propose a session based bit level symmetric key cryptographic technique and it is termed as Spiral Matrix Based Bit Orientation Technique (SMBBOT). SMBBOT consider the input plain text as binary bit stream. During encryption this stream is chopped into manageable sized blocks with variable lengths. Bits of these blocks are taken from MSB to LSB to fit into a square matrix of suitable order following the concept of Spiral matrix. This square matrix splits into 2x2 sub-matrices. Bits are taken column-wise from all 2x2 sub-matrices to form the encrypted binary string. Cipher text is generated from this encrypted binary string. Combination of values of block length and no. of blocks of a session generates the session key for SMBBOT. For decryption the cipher text is considered as binary bit string. Processing the session key information, this binary string is broken down into predefined blocks. Bits of these blocks are taken from MSB to LSB to fit column-wise into 2x2 square matrices. Using these sub-matrices a single square matrix with suitable order is formed. The decrypted binary string is formed after taking the bits from the square matrix following the reverse concept of Spiral Matrix. The plain text is regenerated from decrypted binary string. A comparison of SMBBOT with existing and industrially accepted TDES and AES has been done.*

## 1 INTRODUCTION

Cryptography is an old art of sending secret messages between sender and receiver. With the advancement of internet technologies, cryptography becomes a crucial aspect for secure communications to protect important data from eavesdroppers. Network security is most focused subject which is becoming more and more complex to maintain. As a result researchers are engaged to develop different cryptographic techniques [1, 2, 3, 4, 5, 6, 7, 8, 9, 10] to enhance network security. But each of these techniques has their own merits and demerits. In this paper a new symmetric key cryptographic technique SMBBOT based on session has been proposed. Section 2 of this paper contains the algorithms for encryption, decryption and key generation. Section 3 explains the proposed technique with an suitable example. Section 4 shows the results and comparisons of the proposed SMBBOT with TDES [11] and AES [12]. Section 5 deals with the analysis between above three techniques. Conclusions are drawn in the section 6.

## 2 ALGORITHMS

Three algorithms have been described in this section. First technique explains the encryption technique where as second one explains decryption technique of SMBBOT. Third algorithm explains the concept of session base key.

### 2.1 ENCRYPTION ALGORITHM

It takes source source file / source stream i.e. plain text as input and generates encrypted file / encrypted stream i.e. cipher text as output.

Step1. The plain text is considered as a stream of finite no. of binary bits.

Step2. This binary string breaks into manageable-sized blocks with lengths $(4*n)^2$ like 4 / 16 / 64 / 144 / 256 …… where n = ½, 1, 2, 3, … . First $n_1$ no. of bits is considered as $x_1$ no. of blocks with block length $y_1$ where $n_1 = x_1 * y_1$. Next $n_2$ no. of bits is considered as $x_2$ no. of blocks with block length $y_2$ where $n_2 = x_2 * y_2$ and so on. Finally $n_m$ no. of bits is considered as $x_m$ no. of blocks with block length $y_m$ where $n_m = x_m * y_m$ with $y_m = 8$. So no padding is required.

Step3. Square matrix of order $4*n$ is generated for each block of length $(4*n)^2$. The binary bits of the block are taken from MSB to LSB to fit into this square matrix following the rules of Spiral Matrix along clock-wise direction starting from the cell (1,1).

Step4. Square matrix of order $4*n$ splits into $4*n^2$ no. of 2x2 sub-matrices. Bits are taken column-wise from all 2x2 sub-matrices to form encrypted block with length $(4*n)^2$. The encrypted binary string is formed after taking the bits from all encrypted blocks.

Step5. The cipher text is formed after converting the encrypted binary string into characters.

### 2.2 DECRYPTION ALGORITHM

This algorithm takes encrypted file / encrypted stream i.e. cipher text as input and generates source file / source stream i.e. plain text as output.

Step1. The encrypted file i.e. the cipher text is considered as a binary bit stream.

Step2. After processing the session key information, this binary string breaks into predefined blocks with lengths $(4*n)2$ where n = ½, 1, 2, 3, … .

Step3. Square matrix of order $4*n$ is generated for each block of length $(4*n)^2$. The binary bits of this block are taken from MSB to LSB to fit column-wise into $4*n^2$ no. of 2x2 sub-matrices. Using all these 2x2 sub-matrices a single square matrix of suitable order is generated.

Step4. The decrypted block of length $(4*n)^2$ is generated after taking the bits from the square matrix following the reverse rule of Spiral Matrix. The decrypted binary string is formed after taking the bits from all decrypted blocks.

Step5. The plain text is regenerated after converting the decrypted binary string into characters.

### 2.3 GENERATION OF SESSION BASED KEY

During the encryption process a session based key is generated for one time use in a session of transmission to ensure much more security to SMBBOT. This technique divides the input binary bit stream dynamically into 16 portions, each portion is divided again into x no. of blocks with block length y bits. The final (i.e. $16^{th}$) portion is divided into $x_{16}$ no. of blocks with block length 4 bits (i.e. $y_{16} = 4$). So no padding is required. Total length of the input binary string is given by

$$x_1 * y_1 + x_2 * y_2 + \ldots\ldots + x_{16} * y_{16}.$$

The values of x and y are generated dynamically. The session key contains the sixteen set of values of x and y respectively.

### 3 EXAMPLE

To illustrate the SMBBOT, let us consider a two letter's word "Go". The ASCII values of "G" and "o" are 71 (01000111) and 111 (01101111) respectively. So the binary bit representation of that word "Go" is "0100011101101111" which are taken into a block with length 16 bits [$(4*n)^2$ with n=1]. These bits are taken from MSB to LSB and fit into the square matrix of order 4 [$(4*n)$ with n=1] following the concept of Spiral Matrix along clock-wise direction starting from (1,1) cell. Figure 1 shows the rules of Spiral Matrix. Now this square matrix splits into four [$4*n^2$ with n=1] 2x2 sub-matrices. The encrypted binary string is "0011010011101111" which is formed after taking the bits column-wise from all 2x2 sub-matrices. The equivalent decimal no. of two 8 bit binary numbers 00110100 and 11101111 are 52 and 239 respectively. 52 and 239 are the ASCII values of the characters 4 and ï respectively. So **Go** is encrypted as **4ï**. For decryption process, exactly reverse steps of the above are followed. Figure 2 shows the flow diagram of SMBBOT.

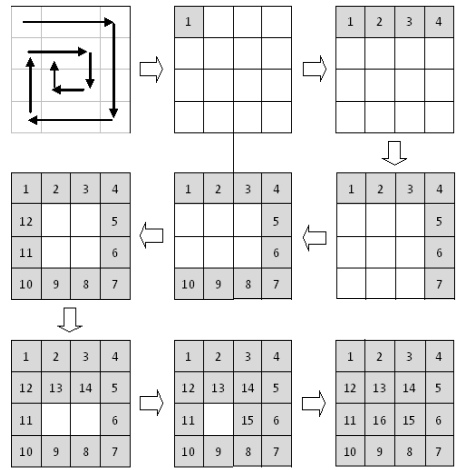

Figure 1: Concept of Spiral Matrix

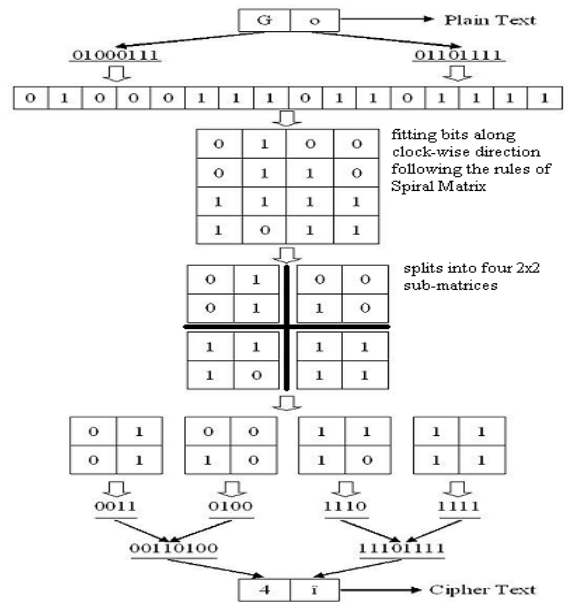

Figure 2: Flow diagram of SMBBOT

### 4 RESULTS AND COMPARISONS

The results includes the comparisons of encryption time, decryption time, character frequencies, Avalanche & Strict avalanche tests and Bit independence values. The comparative study between TDES(168 bits), AES(128 bits) and SMBBOT has done on 20 files of 4 different types with different sizes varying from 882 bytes to 6542640 bytes (6.23 MB).

### 4.1 ENCRYPTION AND DECRYPTION TIMES

Taken time is the difference between processor clock ticks in starting and end. All times are in milliseconds (ms). Table 1 shows the encryption and decryption time for TDES, AES and SMBBOT. Figure 3 and figure 4 show the graphical representation of encryption and decryption times against the file size respectively.

Table 1: Encryption & Decryption times for different files

| Source file size (bytes) | File type | TDES (m.sec.) | | AES (m.sec.) | | SMBBOT (m.sec.) | |
|---|---|---|---|---|---|---|---|
| | | Enc | Dec | Enc | Dec | Enc | Dec |
| 882 | exe | 0 | 15 | 0 | 0 | 0 | 0 |
| 2560 | dll | 0 | 15 | 16 | 0 | 0 | 16 |
| 8181 | txt | 0 | 16 | 62 | 0 | 31 | 15 |
| 12288 | dll | 15 | 16 | 15 | 0 | 16 | 16 |
| 19684 | txt | 16 | 0 | 328 | 0 | 16 | 31 |
| 24064 | doc | 16 | 16 | 0 | 31 | 15 | 0 |
| 55296 | exe | 32 | 16 | 0 | 16 | 31 | 32 |
| 85020 | dll | 32 | 15 | 0 | 32 | 31 | 16 |
| 149848 | txt | 31 | 31 | 31 | 32 | 31 | 31 |
| 351232 | doc | 109 | 78 | 16 | 31 | 78 | 78 |
| 544256 | doc | 125 | 110 | 16 | 46 | 78 | 63 |
| 982732 | txt | 188 | 218 | 78 | 63 | 140 | 141 |
| 1469468 | exe | 297 | 328 | 93 | 78 | 109 | 141 |
| 1925185 | dll | 391 | 500 | 47 | 94 | 218 | 266 |
| 2790946 | txt | 562 | 641 | 187 | 235 | 203 | 265 |
| 3744768 | doc | 750 | 922 | 109 | 172 | 282 | 462 |
| 4377400 | exe | 1062 | 1000 | 172 | 187 | 328 | 297 |
| 5456704 | dll | 1171 | 1157 | 156 | 250 | 475 | 532 |
| 5963500 | txt | 1219 | 1266 | 218 | 329 | 475 | 453 |
| 6542640 | exe | 1375 | 1750 | 375 | 390 | 584 | 392 |

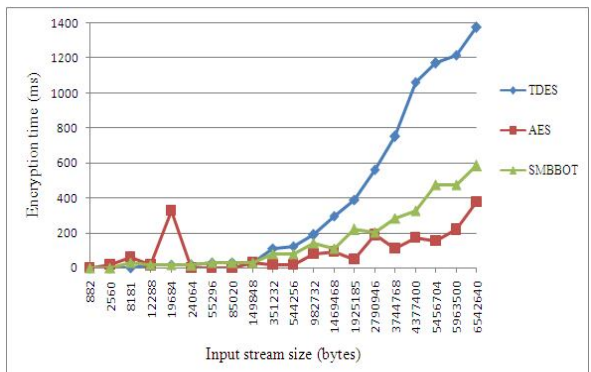

Fig 3: Graphical representation of encryption times v/s file sizes

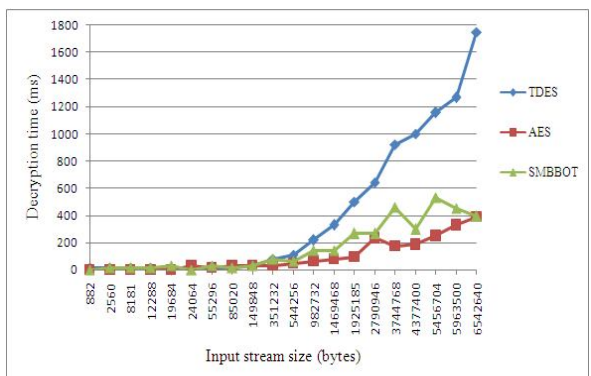

Fig 4: Graphical representation of decryption times v/s file sizes

## 4.2 STUDY OF CHARACTER FREQUENCIES

Study of character frequencies has performed for all twenty files of table 1. Figure 5 shows the distribution of characters in the source file which is text file (file no 19 of table 1 with size 5,963,500 bytes). Figure 6, 7 and 8 show the frequency distribution of characters in cipher text of that text file for TDES, AES and SMBBOT.

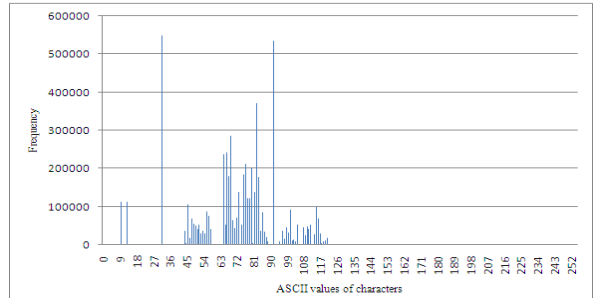

Figure 5: Graphical representation of the frequency distribution of characters for input source stream

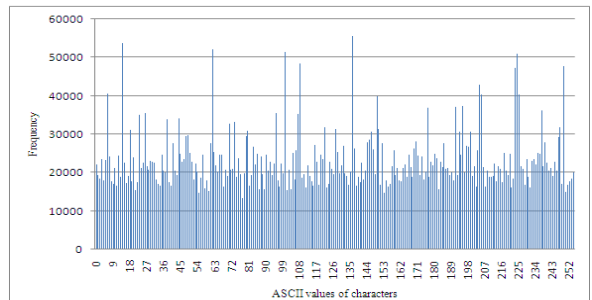

Figure 6: Graphical representation of the frequency distribution of characters for encrypted stream using TDES

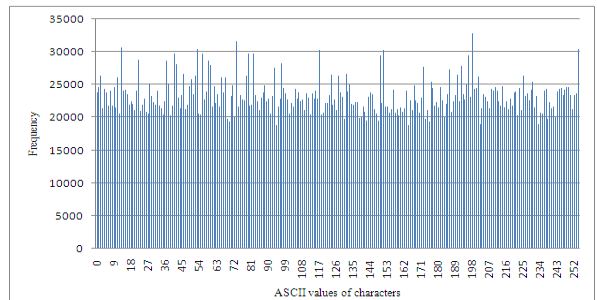

Figure 7: Graphical representation of the frequency distribution of characters for encrypted stream using AES

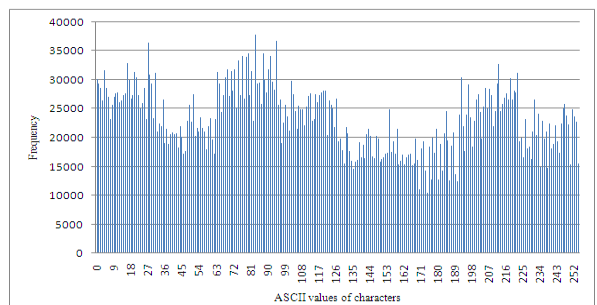

Figure 8: Graphical representation of the frequency distribution of characters for encrypted stream using SMBBOT

## 4.3 TESTS FOR NON-HOMOGENEITY

The test for goodness of fit (Pearson's $\chi^2$) has been performed between the source files (expected) and the encrypted files (observed). The larger Chi-square values (compared with tabulated values) may confirm the higher degree of non-homogeneity between source and encrypted files. Table 2 contains the Chi-square values of twenty different files for TDES, AES and SMBBOT. Figure 9 shows the graphical representation of Chi-square values on logarithmic scale for TDES, AES and SMBBOT.

Table 2: Chi-square values for different files

| Source file size (bytes) | File type | Chi-square values | | |
|---|---|---|---|---|
| | | TDES | AES | SMBBOT |
| 882 | exe | 31047 | 15349 | 7671 |
| 2560 | dll | 36054 | 26036 | 51455 |
| 8181 | txt | 1500874 | 347663 | 796348 |
| 12288 | dll | 165053 | 81474 | 134916 |
| 19684 | txt | 7721661 | 1731310 | 4782854 |
| 24064 | doc | 18918008 | 14182910 | 7424195 |
| 55296 | exe | 252246 | 227972 | 345456 |
| 85020 | dll | 654756 | 433872 | 466220 |
| 149848 | txt | 76621083 | 64043270 | 91356498 |
| 351232 | doc | 537285 | 373599 | 657956 |
| 544256 | doc | 1349490 | 947148 | 1343058 |
| 982732 | txt | 3264221211 | 2585100024 | 4769892987 |
| 1469468 | exe | 3388013 | 3349821 | 4904366 |
| 1925185 | dll | 46245126 | 47627346 | 29721439 |
| 2790946 | txt | 2.426E+10 | 2.197E+10 | 3.075E+10 |
| 3744768 | doc | 7725687 | 5230567 | 3936474 |
| 4377400 | exe | 8783 | 9015 | 10931 |
| 5456704 | dll | 12432371 | 12239623 | 13484571 |
| 5963500 | txt | 10.122E+10 | 8.945E+10 | 12.442E+10 |
| 6542640 | exe | 66742981 | 34387484 | 27636470 |

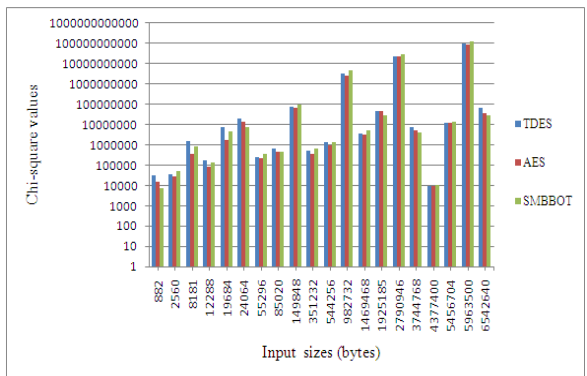

Fig 9: Graphical representation of Chi-square values on logarithmic scale for TDES, AES and SMBBOT

## 4.4 AVALANCHE, STRICT AVALANCHE AND BIT INDEPENDENCE TESTS

Avalanche, Strict avalanche and Bit independence tests has been performed by statistical analysis of data. The bit changes among encrypted bytes for a single bit change in the plain text for the entire or a relative large number of bytes. The Standard Deviation from the expected values is calculated. The ratio of calculated standard deviation with expected value has been subtracted from 1.0 to get the Avalanche and Strict Avalanche values on a 0.0 – 1.0 scale. The value closer to 1.0 indicates the better Avalanche & Strict Avalanche effects and the better Bit Independence criterion. Table 3, 4 and 5 show Avalanche, Strict Avalanche and Bit Independence values respectively. Figure 10, 11 and 12 show the graphical representation of Avalanche, Strict Avalanche and Bit Independence values respectively.

Table 3: Avalanche values for different files

| Source file size (bytes) | File type | Avalanche values | | |
|---|---|---|---|---|
| | | TDES | AES | SMBBOT |
| 882 | exe | 0.9903116 | 0.9955032 | 0.9978863 |
| 2560 | dll | 0.9991916 | 0.9997694 | 0.9946454 |
| 8181 | txt | 0.9993790 | 0.9990676 | 0.9959298 |
| 12288 | dll | 0.9996628 | 0.9994563 | 0.9976442 |
| 19684 | txt | 0.9993587 | 0.9999557 | 0.9908617 |
| 24064 | doc | 0.9991441 | 0.9996181 | 0.9965590 |
| 55296 | exe | 0.9999144 | 0.9999608 | 0.9981813 |
| 85020 | dll | 0.9999630 | 0.9999417 | 0.9969903 |
| 149848 | txt | 0.9998485 | 0.9999418 | 0.9978565 |
| 351232 | doc | 0.9999608 | 0.9999998 | 0.9976464 |
| 544256 | doc | 0.9999880 | 0.9999918 | 0.9968863 |
| 982732 | txt | 0.9999060 | 0.9999283 | 0.9954906 |
| 1469468 | exe | 0.9999968 | 0.9999856 | 0.9981678 |
| 1925185 | dll | 0.9999886 | 0.9999759 | 0.9978458 |
| 2790946 | txt | 0.9999669 | 0.9999855 | 0.9978367 |
| 3744768 | doc | 0.9999958 | 0.9999682 | 0.9983168 |
| 4377400 | exe | 0.9999998 | 0.9999980 | 0.9998145 |
| 5456704 | dll | 0.9999918 | 0.9999949 | 0.9975710 |
| 5963500 | txt | 0.9999840 | 0.9999761 | 0.9944313 |
| 6542640 | exe | 0.9999804 | 0.9999851 | 0.9969527 |

Table 4: Strict Avalanche values for different files

| Source file size (bytes) | File type | Strict Avalanche values | | |
|---|---|---|---|---|
| | | TDES | AES | SMBBOT |
| 882 | exe | 0.9792702 | 0.9838478 | 0.9799445 |
| 2560 | dll | 0.9971642 | 0.9978076 | 0.9910043 |
| 8181 | txt | 0.9972475 | 0.9983127 | 0.9927920 |
| 12288 | dll | 0.9992313 | 0.9993459 | 0.9967422 |
| 19684 | txt | 0.9988550 | 0.9989337 | 0.9875578 |
| 24064 | doc | 0.9984031 | 0.9991557 | 0.9952729 |

| 55296 | exe | 0.9997134 | 0.9997644 | 0.9962980 |
|---|---|---|---|---|
| 85020 | dll | 0.9996336 | 0.9997185 | 0.9962865 |
| 149848 | txt | 0.9994992 | 0.9997208 | 0.9975717 |
| 351232 | doc | 0.9997839 | 0.9997969 | 0.9972040 |
| 544256 | doc | 0.9998611 | 0.9998608 | 0.9944384 |
| 982732 | txt | 0.9996680 | 0.9998524 | 0.9928653 |
| 1469468 | exe | 0.9999451 | 0.9999219 | 0.9974707 |
| 1925185 | dll | 0.9999338 | 0.9999267 | 0.9977467 |
| 2790946 | txt | 0.9999146 | 0.9999552 | 0.9975036 |
| 3744768 | doc | 0.9999493 | 0.9999254 | 0.9976035 |
| 4377400 | exe | 0.9999588 | 0.9999696 | 0.9993187 |
| 5456704 | dll | 0.9999716 | 0.9999750 | 0.9969344 |
| 5963500 | txt | 0.9999090 | 0.9999444 | 0.9903982 |
| 6542640 | exe | 0.9998772 | 0.9999264 | 0.9929675 |

Table 5: Bit Independence values for different files

| Source file size (bytes) | File type | Bit Independence values | | |
|---|---|---|---|---|
| | | TDES | AES | SMBBOT |
| 882 | exe | 0.5416969 | 0.4971214 | 0.4028727 |
| 2560 | dll | 0.5496153 | 0.5545931 | 0.5518652 |
| 8181 | txt | 0.0246086 | 0.0763064 | 0.1783341 |
| 12288 | dll | 0.8042259 | 0.8055426 | 0.8057990 |
| 19684 | txt | 0.0060197 | 0.0326123 | 0.0352535 |
| 24064 | doc | 0.2257191 | 0.0518730 | 0.2212148 |
| 55296 | exe | 0.7475072 | 0.7492728 | 0.7784352 |
| 85020 | dll | 0.8055285 | 0.8146824 | 0.8242885 |
| 149848 | txt | 0.5191341 | 0.5309318 | 0.5489548 |
| 351232 | doc | 0.8310899 | 0.8313195 | 0.8548553 |
| 544256 | doc | 0.6957219 | 0.6982151 | 0.7256481 |
| 982732 | txt | 0.4510866 | 0.4886184 | 0.5436487 |
| 1469468 | exe | 0.6345553 | 0.6288788 | 0.6893819 |
| 1925185 | dll | 0.5884467 | 0.6058651 | 0.6310960 |
| 2790946 | txt | 0.2605189 | 0.2561987 | 0.2503434 |
| 3744768 | doc | 0.9020524 | 0.8937274 | 0.9184160 |
| 4377400 | exe | 0.9981982 | 0.9973744 | 0.9999523 |
| 5456704 | dll | 0.6869081 | 0.6816883 | 0.6934120 |
| 5963500 | txt | 0.3237460 | 0.3274793 | 0.3494796 |
| 6542640 | exe | 0.5776820 | 0.5597410 | 0.5977109 |

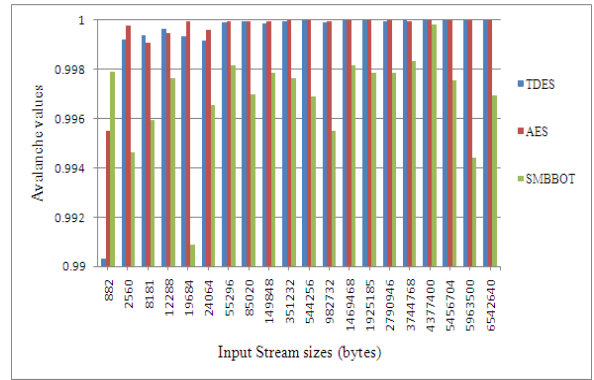

Fig 10: Graphical representation of Avalanche values of different files for TDES, AES and SMBBOT

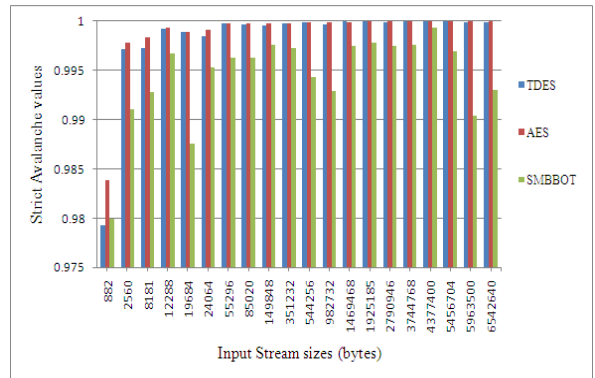

Fig 11: Graphical representation of Strict Avalanche values of different files for TDES, AES and SMBBOT

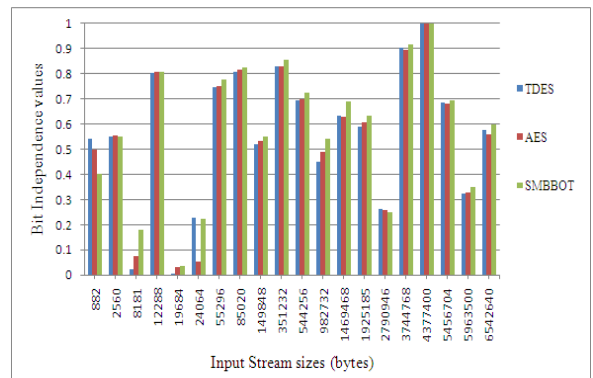

Fig 12: Graphical representation of Bit Independence values of different files for TDES, AES and SMBBOT

## 5  ANALYSIS

Analyzing all the results given in section 4, following are the points obtained after comparing the three techniques TDES, AES and SMBBOT

a) AES takes least times for encryption and decryption compare to other two. For TDES, both the times are 3 times more (approx.) than that of SMBBOT.

b) Figure 3 and 4 show that the encryption and decryption times increases with the increase of input stream sizes.

c) Figure 5 shows that the input bit stream contains some characters with very high and with very low frequencies and some characters with zero frequency. Figure 6, 7 and 8 show that in encrypted files all characters with ASCII values ranging from 0 to 255 are distributed over a certain range. It is harder for a cryptanalyst to detect the original bit stream if the frequency spectrum is smoother. From this point of view, the degree of security of SMBBOT is very much comparable with that of TDES and AES.

d) Figure 9 indicates that the calculated Chi-square values are very high than that the tabulated ones which prove the high degree of non-homogeneity between the source and the encrypted bit stream for all three techniques. So the degree of security of SMBBOT is high and very much comparable with that of TDES and AES.

e) Avalanche, Strict Avalanche and Bit Independence measures are cryptographic test methods which measure the degree of security. Results indicate that the degree of security of SMBBOT is very high and is comparable with that of TDES and AES.

# 6 CONCLUSION

Some of the salient features of SMBBOT are summarized as follows

a) The proposed SMBBOT is very straight forward and simple to understand.

b) The key information varies from session to session for any input bit stream which enhances the security features of the proposed technique.

c) SMBBOT is applicable to ensure high security in message transmission of any form and very much comparable with industry accepted standards TDES and AES.

d) In the proposed session key, the numbers of blocks are taken as power of 3 including 0, which may limit the plain text size to a certain extent. However this value may be changed to any higher integer value i.e. more that 3 to accommodate the plain text of any size.

## REFERENCES:

[ 1]  Sastry V.U.K., Anup kumar K., "A Modified Feistel Cipher involving a key as a multiplicant on both the sides of the Plaintext matrix and supplemented with Mixing Permutation and XOR Operation", International Journal of Computer Technology and Applications (IJCTA), pp. 1-9, JAN-FEB 2012.

[ 2]  Gandhi K.B., Sekhar C.A., lakshmi S.S., "Encryptions of Data Streams using Pauli Spin ½ Matrices and Finite State Machine", International Journal of Computer Applications (IJCA), Vol. 37, No. 2, pp. 8-12, Jan. 2012.

[ 3]  Gupta V., Singh G., Gupta R., "Advance cryptography algorithm for improving data security", International Journal of Advanced research in Computer Science and Software Engineering, Vol. 2, Issue 1, Jan. 2012.

[ 4]  Chowdhury A., Sinha A.K., Dutta S., "Introduction of a Triple Prime Symmetric Key Block Cypher", International Journal of Computer Applications (IJCA), Vol. 39, No. 7, pp. 16-18, Feb. 2012.

[ 5]  Som S., Ghosh S., "A Simple Algebraic Model based Polyalphabetic Substitution Cipher", International Journal of Computer Applications (IJCA), Vol. 39, No. 8, pp. 53-56, Feb. 2012.

[ 6]  Srikantaswamy S.G., Phaneendra, "Enhanced One Time Pad Cipher with More Arithmetic and Logical Operation with Flexible Key Generation Algorithm", International Journal of Network Security & Its Application (IJNSA), Vol. 3, No. 6, pp. 243-248, Nov. 2011.

[ 7]  Dasgupta S., Mazumder S., Paul P., "Implementation of Information Security based on Common Division", International Journal of Computer Science and Network Security (IJCSNS), Vol. 11, No. 2, pp. 51-53, Feb. 2011.

[ 8]  Mazumder S., Dasgupta S., Paul P., "Implementation of Blocked based Data Encryption at Bit-Level", International Journal of Computer Science and Network Security (IJCSNS), Vol. 11, No. 2, pp. 18-23, Feb. 2011.

[ 9]  Nath A., Ghosh S., Mallik M.A., "Symmetric Key Cryptography Using Random Key Generator", Proceedings of the 2010 International Conference on Security & Management, SAM-2010, Vol. 2, pp. 239-244, Las Vegas Nevada (USA), 12-15 July 2010.

[10]  Singhal P., Verma N.K., Laha A., Paul P., Bhattacharjee A.K., "Enhancement of Security through a Cryptographic Algorithm Based on Mathematical Representation of Natural Numbers", International Journal of Computer and Electrical Engineering, Vol. 2, No. 5, Oct. 2010.

[11]  "Triple Data Encryption Standard" FIPS PUB 46-3 Federal Information Processing Standards Publication, Reaffirmed, 1999 October 25 U.S. DEPARTMENT OF COMMERCE/National Institute of Standards and Technology.

[12]  "Advanced Encryption Standard", Federal Information Processing Standards Publication 197, November 26, 2001.